\documentclass[onecolumn]{aa}

\usepackage{graphicx}

\usepackage{txfonts}

\usepackage{natbib}
\bibpunct{(}{)}{;}{a}{}{,}

\begin{document}

\title{Fast electron slowing-down and diffusion in a high temperature
  coronal {X}-ray source}

   \author{R. K. Galloway\inst{1}
          \and
          A. L. MacKinnon\inst{2}
          \and
          E. P. Kontar\inst{3}
          \and
          P. Helander\inst{4}
          }

   \offprints{R. K. Galloway}

   \institute{
             Department of Physics and Astronomy, Kelvin Building, University of Glasgow,
              Glasgow G12~8QQ, U.K.\\
              \email{ross@astro.gla.ac.uk}
              \and
             Department of Adult and Continuing Education, St.\ Andrews
             Building, University of Glasgow, Glasgow G3~6NH, U.K.\\
             \email{a.mackinnon@educ.gla.ac.uk}
             \and
             Department of Physics and Astronomy, Kelvin Building, University of Glasgow,
              Glasgow G12~8QQ, U.K.\\
              \email{eduard@astro.gla.ac.uk}
         \and
             UKAEA, Culham Science Centre, Abingdon OX14~3DB, U.K.\\
             \email{per.helander@ukaea.org.uk}
}

\date{Received 8 October 2004 / Accepted 18 April 2005}

\abstract{
Finite thermal velocity modifications to electron slowing-down rates
may be important for the deduction of solar flare total electron energy.
Here we treat both
slowing-down and velocity diffusion of electrons in the corona at flare
temperatures, for the case of a simple, spatially homogeneous source.
Including velocity diffusion yields a consistent treatment of both
`accelerated' and `thermal' electrons. It also emphasises that one may
not invoke finite thermal velocity target effects on electron lifetimes without
simultaneously treating the contribution to the observed X-ray spectrum
from thermal electrons. We present model calculations of the X-ray
spectra resulting from injection of a power-law energy distribution of
electrons into a source with finite temperature. Reducing the
power-law distribution low-energy cutoff to
 lower and lower energies only increases the relative magnitude
of the thermal component of the spectrum, because the lowest energy
electrons simply join the background thermal distribution. Acceptable
fits to RHESSI flare data are obtained using this model. These also
demonstrate, however, that observed spectra may in consequence be
acceptably consistent with rather a wide range of injected electron
parameters.

}
   
\titlerunning{Fast electron diffusion in a coronal X-ray source}
\authorrunning{R. K. Galloway et al.}

\maketitle

\section{Introduction}
\label{intro}

X- and $\gamma$-ray radiations give the most direct window on accelerated electrons
in flares. They have revealed that accelerated particles, electrons and/or
ions, are an energetically major product of the flare energy release process \citep[e.g.][]{vilmac03}.

\citet{bro:al-03} have emphasised the importance of the source-averaged electron
distribution as a useful `halfway house' between the observed photon spectrum and the 
distribution of electrons initially injected into the source region, i.e.\ the immediate
product of the acceleration process. Assumptions about the dominant factors in electron
transport then allow deduction from the source-averaged electron distribution of the 
distribution output by the acceleration process. Quantities like the total energy
released in the form of fast electrons follow immediately.

\citet{brown71} first analysed the case in which electrons slow down via Coulomb 
collisions in a cold target, i.e.\ a region in which ambient particle thermal speeds 
are all very much less than those of the X-ray emitting electrons. Key results were given for a photon
spectrum $I(\epsilon)$ (photons cm$^{-2}$ keV$^{-1}$ s$^{-1}$) depending on photon energy $\epsilon$ 
as a power-law, i.e.\ $I(\epsilon) \sim \epsilon^{-\gamma}$ for some $\gamma > 0$. Such
a photon spectrum is often observed. Assuming a cold target from which no electrons escape,
it implies an injected electron energy distribution depending on electron
energy $E$ as $E^{-\gamma-1}$. The total energy content of such a distribution is governed 
by the lowest electron energy for which this power-law form holds good. Unfortunately
observations remain ambiguous on the likely value of this lower cutoff, so the total
flare energy in accelerated electrons remains uncertain by more than an order of magnitude.
The flare energy in electrons of energies $>$ 25 keV appears to be a large fraction
of the total flare energy \citep{linhud76, hoy:al-76, saintbenz04}; observations even exist suggesting
a low energy cutoff  in the 2--5 keV range \citep{kan:al-92}.

\citet{ems-03} has pointed out that the cold target assumption may be invalid for the lowest
energy (few keV) accelerated flare electrons. Spatial structure \citep{Emslie_struct} of
RHESSI (Reuven Ramaty High Energy Solar Spectroscopic Imager) images in particular 
suggests that these electrons stop entirely in the corona, in high temperature ($> 10^7$ K)
regions. Then the test particle slowing-down rate no longer increases monotonically with
decreasing particle energy; as particle speed approaches ambient particle thermal speeds from 
above, the rate of loss of energy to background particles decreases, exhibiting a zero 
for a fast particle energy $E_{\mathrm{crit}}$ very close to the ambient electron thermal energy. 
Emslie contends that this value should be used as the minimum possible lower energy cutoff 
when evaluating fast electron total injected energy; electrons below this energy do not slow 
down monotonically, instead merging with the background thermal
distribution.  Emslie's suggested procedure has been employed by \citet{Lin2003} to
estimate the fast electron energy content in the flare of 23rd July 2002.

Emslie's discussion is expressed entirely in terms of the systematic slowing-down rate. This
gives valuable insight but cannot give a complete description of the evolution of injected electrons.
Suppose we inject a mono-energetic electron distribution at $E_{\mathrm{crit}}$. Clearly, although
the \emph{systematic} slowing-down rate at $E_{\mathrm{crit}}$ is zero, the injected electrons will 
not stay indefinitely at $E_{\mathrm{crit}}$; they will spread out in energy in such a way as to 
eventually join the ambient Maxwellian population, doing so in the first instance via
diffusion in velocity rather than systematic slowing down. A more complete treatment is needed
to discuss the form of the photon spectrum and what it is telling us about flare
fast electrons.  \citet{mcclements87} has included velocity
diffusion effects, but only as one component of a complicated
treatment which also features a number of other processes, and he does
not explore the issues we address.

In this contribution we examine the evolution of injected electrons when the cold target
assumption breaks down, in the slightly idealised case of a homogeneous source and constant
background temperature. We include velocity diffusion as well as systematic slowing down,
and reformulate the interpretation of observed photon spectra. The next Section formulates 
the problem and gives some analytical discussion. Section~\ref{numer} illustrates our 
discussion with some calculated spectra, compared to RHESSI data. Section~\ref{end}
gives brief conclusions.

\section{Velocity diffusion of fast electrons}
\label{fp}

\subsection{Assumptions; Fokker-Planck equation}

In order to illustrate the consequences of velocity diffusion for photon spectra we consider 
an idealised problem in which all injected electrons thermalise in a uniform, homogeneous medium,
characterised by a single, ambient electron density $n_e$ and
temperature $T_e$. Loss at boundaries will have a negligible influence on the electron distribution function and pitch-angle
information will not be important for the calculation of the total, emergent photon spectrum,
particularly since bremsstrahlung directionality is unimportant at the few keV photon energies appropriate here. 
We can gain significant insight, and also solve a problem appropriate to understanding the X-ray 
emission integrated over the whole of the event, by studying a steady-state situation, so no quantity
depends on time. In practice $n_e$ and $T_e$ will evolve as a result of the thermal and hydrodynamic
response of the atmosphere to the flare energy release, but this is a complication of detail rather 
than principle and we ignore it in the interests of gaining insight.
 (Relevant electron timescales such as the thermalisation time are at
most on the order of seconds, whereas bulk changes
to the plasma, such as changes of temperature as characterised by
variation of the soft {X}-ray flux, take place over timescales of a
few minutes.) Thus we can characterise the 
(pitch-angle integrated) electron distribution everywhere in the source by a single function $f(v)$ 
((cm s$^{-1}$)$^{-3}$) of velocity $v$ (cm s$^{-1}$). The normalisation of $f$ is given by

\begin{equation}
\label{normal}
4 \pi \int_0^\infty f(v) v^2 \, \mathrm{d}v \, = \, N_e \, \mbox{,}
\end{equation}

\noindent
where $N_e$ is the total number of all electrons in the source.  

An electron of 10 keV initial energy will stop in a column depth of $2
\times 10^{19}$ cm$^{-2}$ of fully ionised hydrogen, inside the coronal
portion of a loop, e.g.\ for densities $> 2 \times 10^{10}$ cm$^{-3}$ and
loop lengths $> 10^9$ cm, conditions not infrequently inferred in flares.
Thermalisation, in the alternative case that this energy is close to the
thermal speed, will occur in a comparable distance. In addition, magnetic field
convergence may further enhance the coronal residence time
of electrons and increase the effective distance available for thermalisation.  Electrons well above
thermal speeds will experience cold target conditions throughout the
corona and chromosphere and will in any case be described correctly by
what follows. Thus, while possibly not the case in all events, it is not
unreasonable that all electrons for which finite thermal velocity (`warm') target effects are
important thermalise in the coronal, warm target region.

This
steady-state treatment will be valid as long as we use it on times that
are not long enough for the injected electrons to become significant in
number compared to the thermal distribution, but are longer than relaxation
times for most injected electrons. Alternatively, we may regard it as
giving the time integral of the distribution function in the case of an
initial, impulsive injection, in which case the source function $S$ is
actually the initial condition on $f$ \citep{mac:cra-91}. The time
integral of $f$ is the necessary quantity for calculation of the total
bremsstrahlung photon yield.

We use the Fokker-Planck formalism for particle evolution under binary collisions \citep[e.g.][]{ros:al-57,
mon:tid-64}.  We also make the assumption that the fast particles 
are `dilute', in the particular sense that they may be ignored in calculating the velocity space drift 
and diffusion coefficients:  these may be evaluated purely from the background distribution. Then the steady-state,
Fokker-Planck equation for $f(v)$, derived from
\citet[][p.37--38]{hel_sig}, may be written as

\begin{equation}
\label{fpeqn}
- \frac{1}{v^2} \frac{\partial}{\partial v}\left(
  \left\{ \Phi \left( v \right) - v \,
      \Phi^\prime \left( v \right) \right\} \left\{f +
      \frac{1}{2v}\frac{\partial f}{\partial v}\right\} \right )
  \, = \, S(v) \, \mbox{,}
\end{equation}

\noindent
where the function $\Phi$ is the error function:

\begin{equation}
\Phi(x) \, \equiv \, \mathrm{erf}(x) \, = \, \frac{2}{\sqrt{\pi}} \int^x_0 \mathrm{e}^{-y^2}
{\,\mathrm{d}} y \, \mbox{.}
\end{equation}
 
Here velocities $v$ have been normalised to the ambient electron
thermal speed $v_T = \sqrt{2kT/m_e}$, and times to the electron
thermal collision time $t_c = 4 \pi \epsilon_0^2 m_e^2 v_T^3
n_e^{-1} e^{-4} (\ln \Lambda)^{-1}$
Although there is no time-dependence in this problem, the electron injection function $S(v)$ is of course per unit time.

Equation (\ref{fpeqn}) needs two boundary conditions. The boundary
condition at infinity is: $(f + 1/2v \, \partial f / \partial v) \to
0$ as $v \to \infty$.  This condition ensures that there is no flux of
particles out of the system at infinity.
For the other boundary condition we fix $f(v)$ at $v = 0$: $f(0) = f_0$,
consistent with the conditions for validity of our steady-state
treatment and with the assumption of `diluteness' that justified the
linearisation of the Fokker-Planck equation. $f_0$ describes the
background thermal distribution. We integrate Eq.~(\ref{fpeqn}) once from $v$ to infinity and use the boundary
condition at infinity, then again from $0$ to $v$, employing an
integrating factor $\mathrm{e}^{v^2}$ and using the boundary
condition at $0$.  Thus we find the solution

\begin{equation}
\label{fpsol}
  f(v) \, = \, f_0 \, \mathrm{e}^{-v^2} + 2 \int^v_0 \frac{v^{\prime} \mathrm{e}^{\left(
        v^{\prime 2} - v^2 \right)}}{
    \Phi(v^{\prime}) -
    v^{\prime} \Phi^{\prime}(v^{\prime}) } {\,\mathrm{d}} v^{\prime}
  \int^{\infty}_{v^{\prime}} v^{\prime \prime 2} S(v^{\prime \prime})
  {\,\mathrm{d}} v^{\prime \prime} \, \mbox{,}
\end{equation}

\noindent
which will be used in Sect.~\ref{numer} to calculate distributions and resultant photon spectra for various forms of 
$S$. The result~\ref{fpsol} is rendered rather impenetrable, however, by the function $\Phi$.
An illuminating, semi-quantitative analytical discussion may be carried out by invoking the large argument form
of $\Phi(v)$, strictly applicable for $v \gg 1$, in which $\Phi(v)$
approaches unity for large $v$. In this case the Fokker-Planck equation becomes:

\begin{equation}
\label{fpapp}
-\frac{1}{v^2}\frac{\partial}{\partial v}\left(f +
  \frac{1}{2v}\frac{\partial f}{\partial v}\right)\,=\,S(v) \, \mbox{.}
\end{equation}

\subsection{Approximate solution}

First, we note that the LHS of Eq.\ (\ref{fpapp}) may be rewritten

\begin{displaymath}
-\frac{1}{v^2}\left(1-\frac{1}{2 v^2}\right)\frac{\partial f}{\partial
  v} - \frac{1}{2 v^3}\frac{\partial^2 f}{\partial v^2} \, \mbox{,}
\end{displaymath}

\noindent
showing explicitly that the systematic rate of change of $v$ (the
coefficient of $\partial f / \partial v$)
does indeed display a zero, even in this approximate form, at $v=1/\sqrt{2}$, quite close to the zero found using the full form of 
$\Phi$ by \citet{ems-03}. All the necessary qualitative features are included in the description
of Eq.\ (\ref{fpapp}), in spite of its approximate nature, and its solutions will have the appropriate qualitative 
properties.

Note that the warm target corrections to the systematic slowing-down rate
and the dispersive term both become important in the limit $v \to 1/\sqrt{2}$.
The
arguments of \citet{ems-03} rest on the presence of the zero (here at $v =
1/\sqrt{2}$) in the
systematic slowing-down term. Electron slowing-down times approach
$\infty$, so a finite
emergent photon spectrum demands $S \to 0$ as $v \to 1/\sqrt{2}$.
However, in this limit the dispersive term has become important,
removing the divergence in electron `lifetime'.
Using the boundary conditions as before, and changing the order
  of integration in the resulting integral, Eq.\ (\ref{fpapp}) has
the solution

\begin{equation}
\label{appsol}
f(v) \, = \, f_0 \mathrm{e}^{-v^2} \, + \,
\mathrm{e}^{-v^2}\int_0^v u^2 S(u) \left(
  \mathrm{e}^{u^2}-1\right) \,\mathrm{d}u \, + \, \left(1 -
  \mathrm{e}^{-v^2}\right) \int_v^\infty u^2 S(u) \, \mathrm{d}u \, \mbox{.}
\end{equation}

In the absence of any source $S$, Eq.\ (\ref{fpapp}) has the background Maxwell-Boltzmann distribution ($f_0 \mathrm{e}^{-v^2}$) 
as its solution, as indeed does Eq.\ (\ref{fpeqn}). The description in terms only
of systematic slowing-down rates divorces the fast particle and background distributions. 
This is no longer the case in this diffusive treatment: the presence of the boundary condition 
at $v = 0$, which must be satisfied using the same background density used to calculate
the drift and diffusion coefficients, ensures that the fast particle distribution merges 
smoothly with the thermal `core'. It follows that we are obliged to include also 
the contribution to photon emission from the thermal plasma, if we are indeed looking at 
photon energies such that velocity diffusion is important for the emitting electrons.

In the limit $v \to \infty$, Eq.\ (\ref{appsol}) becomes

\begin{equation}
\label{vinf}
f(v) \, \to \,  \int_v^\infty u^2 S(u) \, \mathrm{d}u \, \mbox{.}
\end{equation}

\noindent
Recall that our $f$ is identical with the mean source electron
distribution, the key quantity in interpreting X-ray emission \citep{bro:al-03}. With the traditional assumptions of fast electrons slowing
down in a cold, thick target, this mean distribution is just the
cumulative distribution of the injected energy distribution. Eq.~(\ref{vinf})
reproduces this result, as it should. We can rapidly recover well-known results
in that limit, for instance Brown's (1971) relations between energy
power-law spectral indices of observed photons and injected electrons.

We see that the three terms in the solution (\ref{appsol}) consist of: the Maxwell-Boltzmann core of the distribution;
a term which resembles the cold target result more and more closely as $v \to \infty$; and a term which forces
these two regimes to merge smoothly.

\subsection{Mono-energetic injected population}
\label{monoan}
The special case of a mono-energetic form for $S$ is instructive:

\begin{equation}
\label{deltas}
S(v) \, = \frac{S_0 \delta(v-v_0)}{v^2} \, \mbox{,}
\end{equation}

\noindent
for some velocity $v_0$.
Then the solution (\ref{appsol}) becomes:

\begin{equation}
f(v)\, = \, 
\cases{
  \left\{f_0 + S_0\left(\mathrm{e}^{v_0^2}-1\right)\right\}
  \mathrm{e}^{-v^2} & if $v > v_0$\,, \cr
  f_0 \mathrm{e}^{-v^2} + S_0 \left(1 -
      \mathrm{e}^{-v^2}\right) & if $v < v_0$\,.
    }
    \label{del}
\end{equation} 

\begin{figure}
  \centering
  \resizebox{\hsize}{!}{\includegraphics{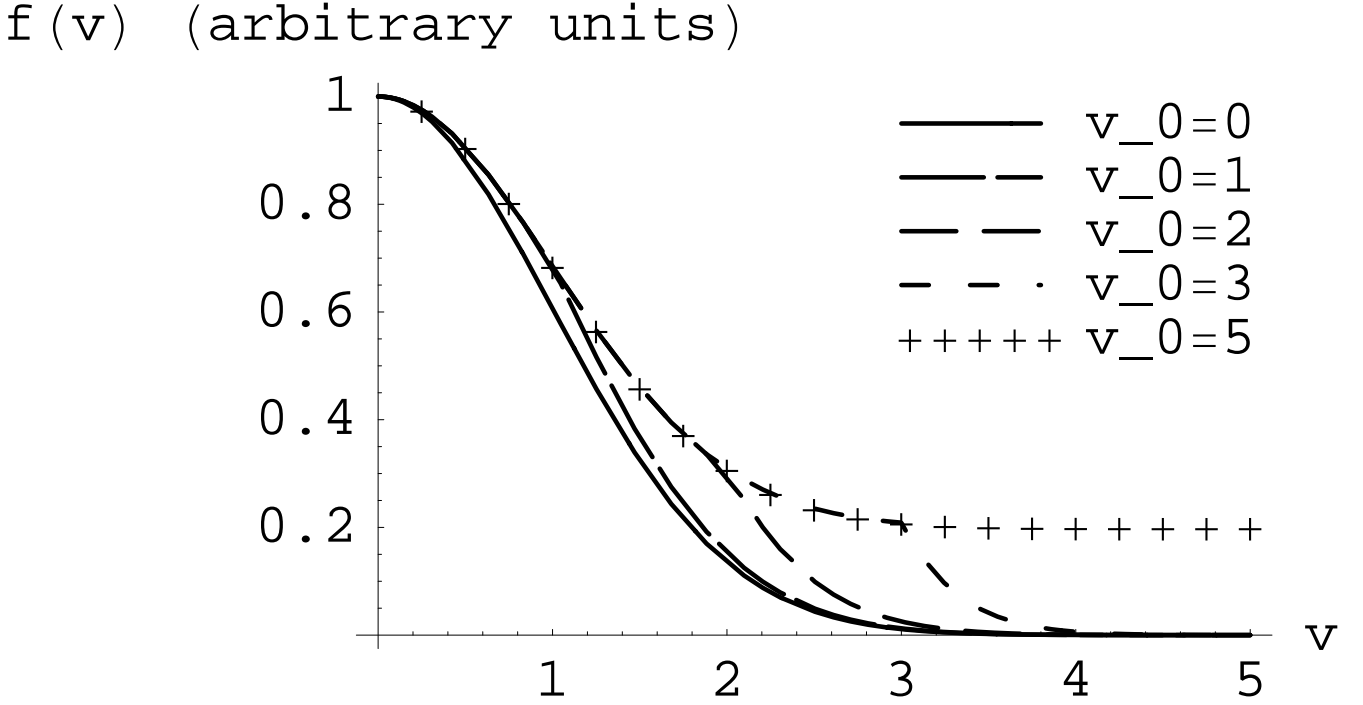}}
  \caption{Form of the relaxed, steady-state distribution resulting from 
injected, monoenergetic electrons. The different curves represent different
values of the velocity (in units of the thermal speed) of the injected
electrons.  The noticeable change in behaviour between the curves for
$v_0=2$ and $v_0=3$ highlights the
injected electrons' transition from being in the thermalisation regime
to being properly non-thermal.}
\label{deltafig}
\end{figure}

\noindent
For $v < v_0$ the distribution is composed of the original, background Maxwellian distribution
plus a component which is identical to the cold target result for $v, v_0 \gg 1$, but 
which approaches $0$ as $v \to 0$. This additional, non-Maxwellian 
component becomes less and less significant for smaller and smaller $v_0$. For $v > v_0$, 
the distribution is identical with the original background Maxwell-Boltzmann distribution, 
only with its normalisation increased. Since we must have $S_0 \ll f_0$ for 
validity of the original linearisation of the Fokker-Planck equation, we see that the 
distribution will resemble the original Maxwellian more and more closely as $v_0$ gets 
closer and closer to 0. This justifies the qualitative comments made in Sect.~\ref{intro}: 
electrons injected close to the thermal speed diffuse in energy 
rather than slowing down monotonically, merely adding their number to the original 
background Maxwellian.  Figure \ref{deltafig} illustrates this. 

\subsection{Deduction of $S$ for a power-law photon spectrum}

As mentioned in Sect.~\ref{intro}, the photon spectrum may in principle be inverted to yield the
mean (source-averaged) electron distribution \citep{brown71,bro:al-03}. This is identical with the 
distribution function $f$ in the special case of our homogeneous source. If observations have given 
us $f$ in this way, Eqs.\ (\ref{fpeqn}) or (\ref{fpapp}) then immediately give $S$. Consider the case 
of a power-law photon spectrum $I(\epsilon) \sim \epsilon^{-\gamma}$. Assume for the moment that 
this form holds at all photon energies of interest. Then the results of \citet{brown71} give us 
$f(v) \sim v^{-2 \gamma -3}$. Inserting this form into Eq.~(\ref{fpapp}) we find

\begin{equation}
\label{powapp}
S(v) \, \sim \, (2 \gamma + 3)\, v^{- 2 \gamma - 6} \left\{1 -
  \frac{\left(2 \gamma +
  5 \right)}{2} \frac{1}{v^2} \right\} \, \mbox{.}
\end{equation}

As in the case retaining only systematic energy loss \citep{ems-03}, $S$ has a zero, changing sign at

\begin{displaymath}
v_* \, = \, \sqrt{\gamma + 5/2} \, \mbox{.}
\end{displaymath}

\noindent
$S(v)$ takes negative values for $v < v_*$. We have seen that a diffusive treatment underlines the necessity of including
the radiation from the thermal, `core' part of the distribution. In assuming the power-law photon spectrum to be
appropriate at all photon energies we have implicitly neglected this contribution. 
 $v_*$ represents the lowest velocity
at which the single, uninterrupted power-law photon spectrum can still be reconciled with the presence of the
Maxwell-Boltzmann core. Below $v_*$ we would have to actually remove particles from this core to prevent
deviations from a power-law photon spectrum; hence $v_*$'s dependence
on $\gamma$. Moreover, we cannot `overcome' the core Maxwellian
distribution by, for example, injecting
a power-law energy distribution of electrons that persists down in energy
towards thermal speeds. As we saw in Sect.~\ref{monoan}, these electrons
mostly thermalise diffusively, producing only a slight modification to the core.

We might follow \citet{ems-03} and 
evaluate total electron energy content by 
integrating $S$, given by Eq.\ (\ref{powapp}), from $v_*$ to $\infty$. Rather than adopting a 
lower energy cutoff for the power-law which evidently holds at high energies, however, this 
approach underlines the need for a consistent treatment of radiation from
both thermal and accelerated electrons. 

\section{Numerical illustrations and application to data}
\label{numer}
\subsection{Numerical illustrations}
We return now to the full solution of the Fokker-Planck equation as
given in Eq.\ (\ref{fpsol}), and provide some illustrative examples relevant to solar
observations.  We adopt as the source function $S(v)$ a
power-law, employing the form
\begin{eqnarray}
\label{Seq}
S(v) & = & \tilde{S}(v) \, H(v-v_0) \nonumber \\
     & = & S_0 \,
       (\delta_v-3) \, v_0^{\delta_v -3} \, v^{-\delta_v} \, H(v-v_0) \, \mbox{,}
\end{eqnarray}

\noindent
where $\tilde{S}$ is normalised such that per unit time (normalised to
the electron thermal collision time) there are $S_0$ particles
injected in total at velocities above $v_0$, and $S$ is prevented from
going to infinity at low velocities by Heaviside's step
function $H$ which removes all particles with velocities less than
$v_0$.

For a homogeneous source, the emission rate of photons of
  energy $\epsilon$ per unit energy range per unit volume, ${\,\mathrm{d}} j /
  {\,\mathrm{d}} \epsilon$ (photons s$^{-1}$ keV$^{-1}$ cm$^{-3}$),
  may be found by multiplying the distribution function by $v\,
    \mathrm{d}\sigma/\mathrm{d}\epsilon$ to obtain the instantaneous rate of emission of
    photons by electrons in the velocity range ${\bf v} \rightarrow
    {\bf v} + {\bf d^3v}$, then integrating over all velocities \citep{brown71}, noting that
    ${\bf d^3v}~=~4 \pi v^2 {\bf dv}$.  This gives

 \begin{equation}
   \label{bremeq}
  \frac{{\,\mathrm{d}} j}{{\,\mathrm{d}} \epsilon} = n_e v_T^3 \int^{\infty}_{\sqrt{
        \epsilon / kT}} f(v) \frac{{\,\mathrm{d}} \sigma}{{\,\mathrm{d}} \epsilon} v^3
  {\,\mathrm{d}} v \, \mbox{,}
  \end{equation}

\noindent
  where $n_e$ is the background plasma number density and $\frac{{\,\mathrm{d}}
    \sigma}{{\,\mathrm{d}} \epsilon}$ is the Bethe-Heitler
  cross-section:

  \[
  \frac{{\,\mathrm{d}} \sigma}{{\,\mathrm{d}} \epsilon} \, = \, \frac{Q_0 m_e
    c^2}{\epsilon E} \ln
  \left(\frac{1+\sqrt{1-\epsilon/E}}{1-\sqrt{1-\epsilon/E}} \right) \, \mbox{.}
  \]

\noindent
Here, $E$ is the electron kinetic energy and $Q_0$ is given by

\[
Q_0 \, = \, \frac{8}{3} \alpha r_e^2 \, \mbox{,}
\]

\noindent
where the fine structure constant $\alpha \approx 1/137$ and
$r_e=2.82\times10^{-13}\,\mathrm{cm}$ is the classical electron
radius.  The photon spectrum, ${\,\mathrm{d}} \tilde{j} /
  {\,\mathrm{d}} \epsilon$ (photons s$^{-1}$ cm$^{-2}$ keV$^{-1}$),   that would be observed by RHESSI is given
by

\begin{equation}
\frac{{\,\mathrm{d}} \tilde{j}}{{\,\mathrm{d}} \epsilon} \, = \, \frac{V}{4 \pi r^2_{\oplus}} \frac{{\,\mathrm{d}} j}{{\,\mathrm{d}}
  \epsilon} \, \mbox{,}
\label{speceq}
\end{equation}

\noindent
where $V$ is the volume of the source and $r_{\oplus}$ is the
distance from the Sun to the Earth.  Since radiation from the
whole emitting volume is observed, the value of $V$ will be determined
implicitly by the spectral fitting process (see Sect.\ \ref{data})
and need not be separately evaluated.

As previously stated, the presence of $\Phi$ in Eq.\ (\ref{fpsol}) 
renders a full analytical solution intractable. Therefore we proceed
numerically, using Romberg integration \citep{num-rec} to evaluate
Eq.\ (\ref{bremeq}) with $f$ given by Eq.\ (\ref{fpsol}).

The parameters which may be varied in the numerical simulations are:
the ratio of $f_0$ to $S_0$ i.e.\ the relative magnitudes of the
background Maxwellian population and the injected power-law electrons;
the lower cutoff velocity of the injected electrons, $v_0$; and the
spectral index $\delta$ of the power-law.  It should be noted that in
the majority of the literature, the power-law used to model flare
electrons is a power-law in energy of the form $S(E)=S_0 E^{-\delta}$.
For consistency and ease of comparison we shall refer to this $\delta$
in subsequent discussion.  The corresponding $\delta_v$ for our
velocity power-law, Eq.\ (\ref{Seq}), is related to $\delta$ by the
expression $\delta_v=2\delta-1$.  Unless otherwise stated, the default
values of the parameters are $f_0/S_0=10^8$, $\delta=4.0$, and $v_0=v_T$.

\begin{figure}
\centering
\resizebox{\hsize}{!}{\rotatebox{270}{\includegraphics{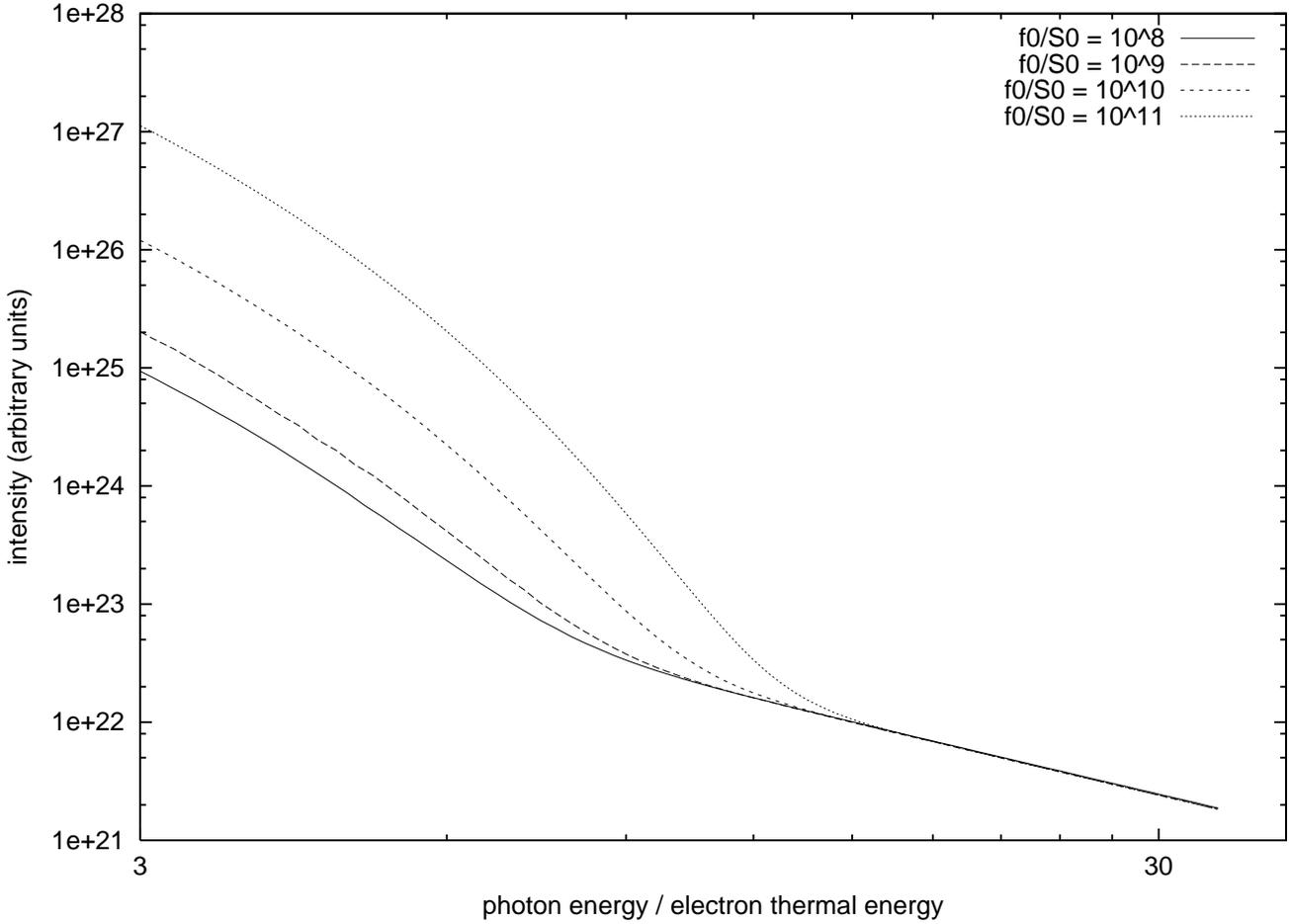}}}
\caption{Simulated X-ray photon spectra, showing the effect of
  altering the ratio of $f_0$ to $S_0$.  The photon energy is
  normalised to the kinetic energy of an electron of thermal velocity.
   The emitted intensity values are arbitrary, since we are interested
   only in relative changes to the spectral profiles.}
\label{f0plot}
\end{figure}

Figure \ref{f0plot} illustrates the effect of altering the ratio
$f_0/S_0$.  The logarithmically-plotted photon spectra consist of two main regions:  a
straight power-law profile at high photon energy, blending smoothly
into a Maxwellian profile at lower energy.  As would be expected,
increasing the relative contribution of the Maxwellian background has
no effect at high photon energies since here the profile only contains
contributions from electrons of the photon energy or higher.  However,
a larger $f_0/S_0$ value leads to a correspondingly higher
contribution to the Maxwellian portion of the spectra from the
background plasma.  Furthermore, this larger value also corresponds to
an increase in the photon energy up to which the Maxwellian impinges on
the otherwise straight power-law: the profile departs from the
straight portion at higher energy for a larger $f_0/S_0$.

\begin{figure}
\centering
\resizebox{\hsize}{!}{\rotatebox{270}{\includegraphics{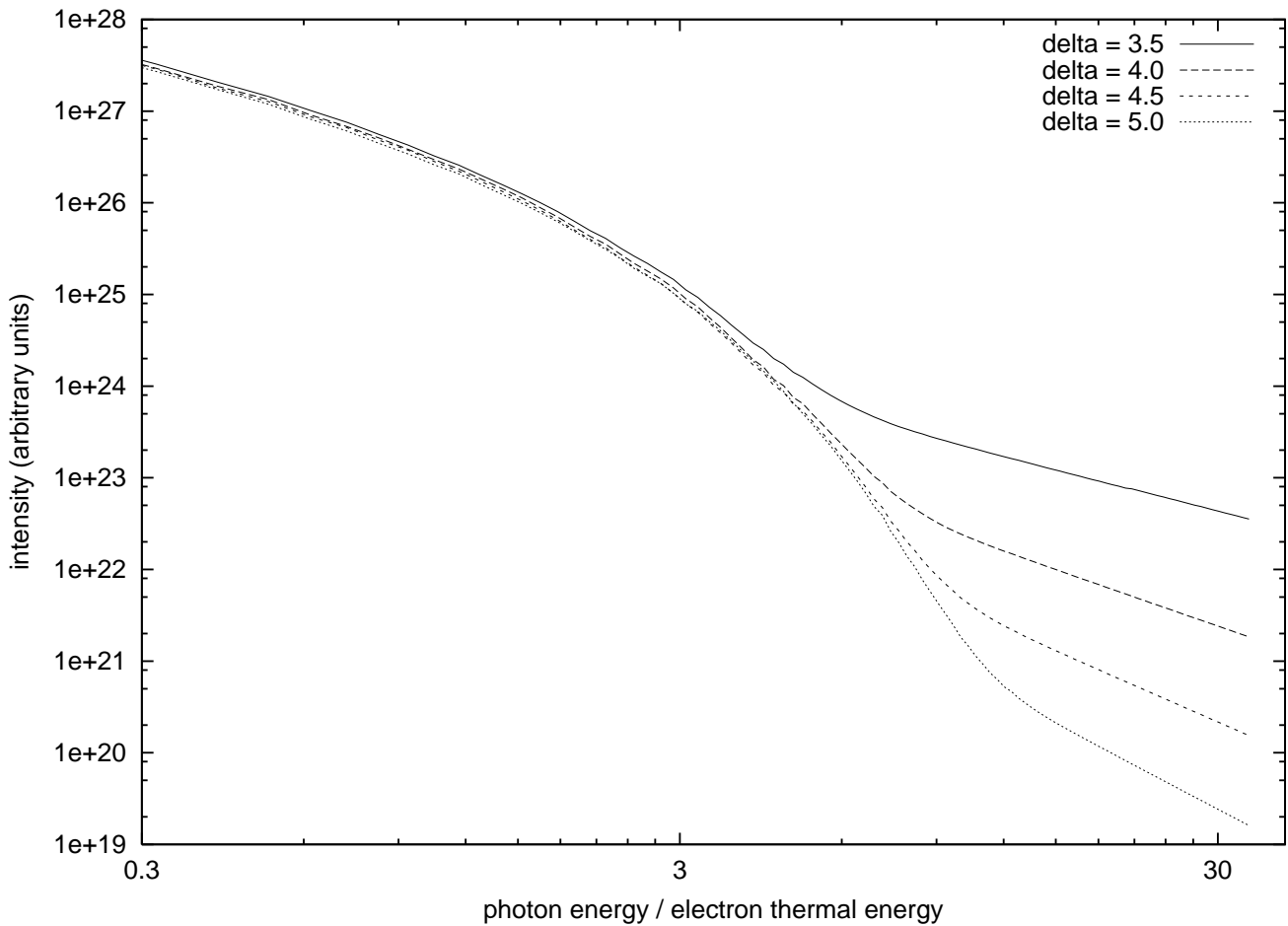}}}
\caption{As Fig.\ \ref{f0plot}, but with variation of the power-law
  spectral index $\delta$, and over an expanded range of photon energy.}
\label{deltaplot}
\end{figure}

The alteration of the electron energy spectral index $\delta$ is
depicted in Fig.\ \ref{deltaplot}.  As may be seen, this has minimal
effect at low photon energy, but a larger $\delta$ results in a
correspondingly steeper slope in the power-law region of the spectrum.
 A greater value for $\delta$ also causes a more rapid reduction
 in the total number of electrons as a function of increasing energy
 in the injected population.  Thus, a higher $\delta$ leads to a
 relative reduction in the intensity of the power-law spectrum for a
 given photon energy.  This reduced contribution from the power-law
 electrons also increases the photon energy up to which the Maxwellian
 element forms a significant part of the resultant profile.
 Consequently, the departure from the straight power-law profile
 occurs at a higher photon energy for larger $\delta$ values, visually
 mimicking a non-existent change in the power-law low energy cutoff.

\begin{figure}
\centering
\resizebox{\hsize}{!}{\rotatebox{270}{\includegraphics{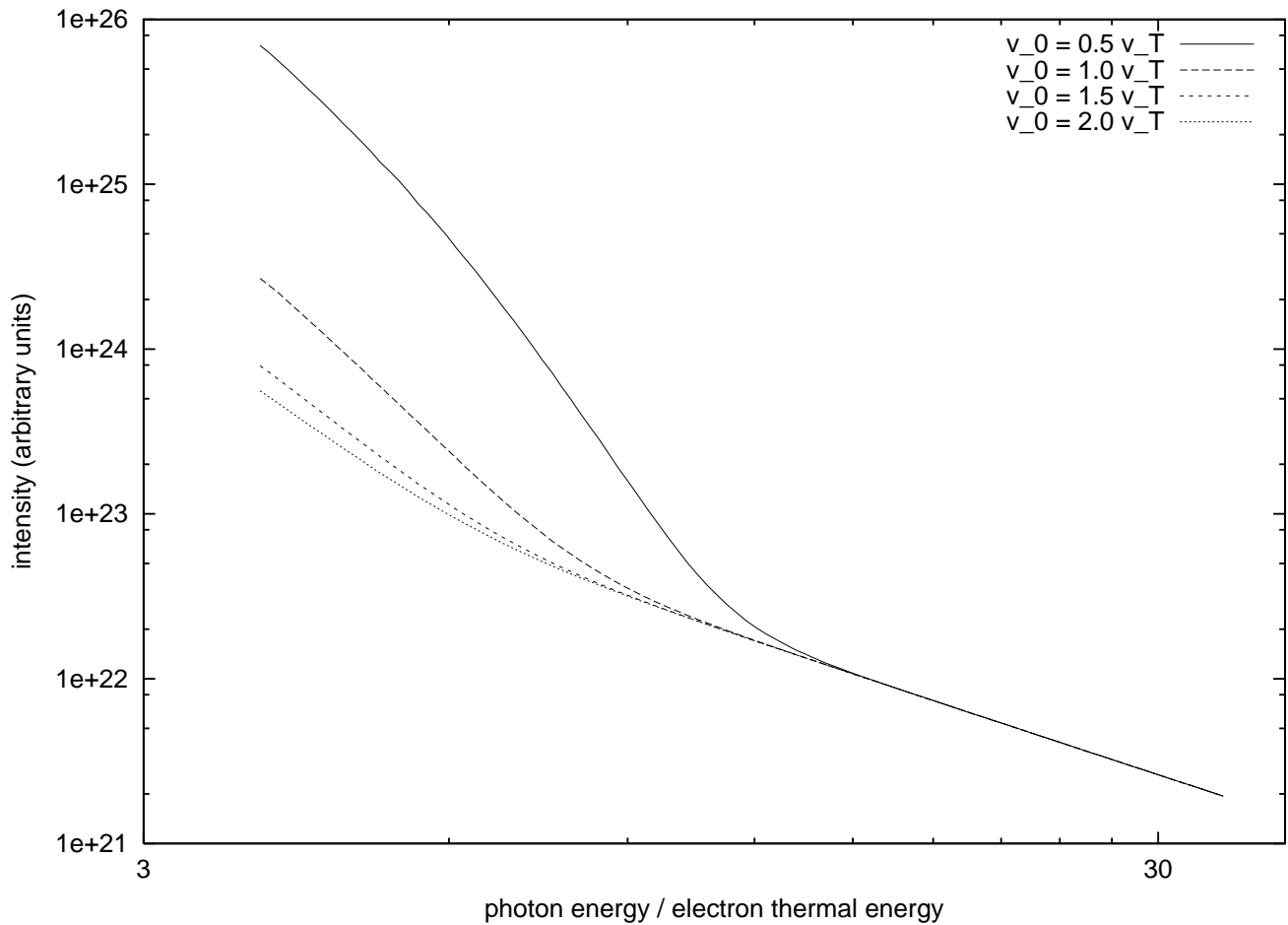}}}
\caption{As Fig.\ \ref{f0plot}, but with variation of the power-law cutoff $v_0$.}
\label{v0plot}
\end{figure}

Actual variation of the cutoff, $v_0$, is shown in Fig.\ \ref{v0plot}.
Since electrons injected below a few times $v_T$ thermalise rather
than slowing down systematically, allowing the cutoff to extend to lower
energies merely adds electrons to the `background' Maxwell-Boltzmann
distribution. This explains the counterintuitive result, clearly visible in Fig.\ \ref{v0plot}, that a \emph{lower} value of
$v_0$ results in a spectrum which attains power-law form at
\emph{higher} photon energies:  the large number of injected electrons
at low energies thermalise and enhance the Maxwell-Boltzmann
distribution, concealing the lower-energy portion of the power-law form.

\subsection{Comparisons to RHESSI data}
\label{data}

\begin{figure}
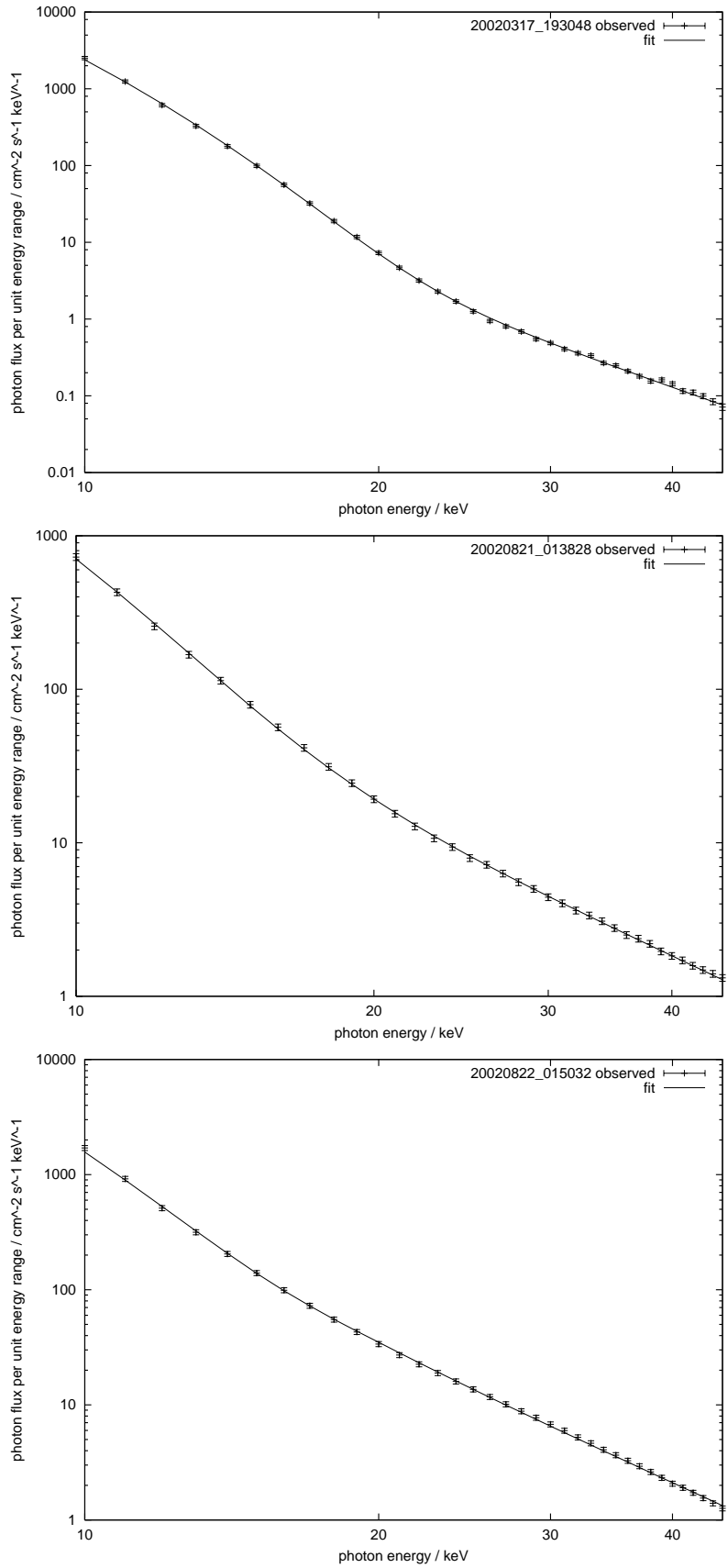

\centering
\resizebox{0.6\hsize}{!}{\rotatebox{270}{\includegraphics{2137fig5a.eps}}}
\resizebox{0.6\hsize}{!}{\rotatebox{270}{\includegraphics{2137fig5b.eps}}}
\resizebox{0.6\hsize}{!}{\rotatebox{270}{\includegraphics{2137fig5c.eps}}}
\caption{X-ray photon spectra recorded by RHESSI during
  the flares of 17th March 2002 at 19:30:48 UT, 21st August 2002 at 01:38:28 UT,
  and 22nd August 2002 at 01:50:32 UT.   Model
  fits to the
  observed spectra are shown as solid lines.}
\label{flareplots}
\end{figure}

The recent launch of the Reuven Ramaty High Energy Solar Spectroscopic Imager 
(RHESSI) has opened a new era in
high resolution X-ray spectroscopy of solar flares \citep{Lin2002}. The 
analysis of solar flare
spectra has revealed statistically significant deviations from a simple 
isothermal and power-law model \citep{Kontar2003}. The detailed analysis of the X-ray producing 
spectra using a model-independent inversion technique \citep{Piana2003} shows deviations from the
pure isothermal model.
These new findings can be treated  as the manifestation of velocity 
space diffusion in warm
target plasma.

For illustrative purposes, we consider a few example events with sufficiently high count rate to provide 
reliable photon statistics. We have
limited our analysis to within the energy range 10--50 keV, where thermal and 
nonthermal components
merge. Below 9 keV, the bremsstrahlung continuum is contaminated by a 
complex of strong iron
lines. Above $\sim$ 50 keV, spectral features not related to the model 
discussed become
dominant \citep{Kontar2003}.

  We fit the model spectra to the observed spectra by optimising over 4 
parameters:  the relative
magnitudes of the background and injected electron populations, $f_0/S_0$; 
the injected electron
power-law low-energy cutoff, $v_0$; the background temperature, $T_e$; and  
the injected electron
power-law spectral index, $\delta$.  The optimisation seeks to
minimise an un-normalised $\chi^2$
 fit statistic -- in this case absolute values of $\chi^2$ must
 be treated with caution since the process of
 deconvolving the RHESSI instrument response from the observed counts
 spectrum to produce the photon spectrum introduces an element of
 error on each photon spectrum data point which is
 difficult to quantify precisely \citep{smith_2002}.
However, we only compare relative values of the fit statistic to optimise
the model fits, so un-normalised $\chi^2$ is sufficient for our purpose.  
As may be seen
from Fig.\ \ref{flareplots}, sets of optimal model parameters may be
found which give model spectra that closely match the observed RHESSI spectra.
 The model parameters corresponding to the smallest
fit statistic in each case are given in Table \ref{paramstab}.

\begin{table}
\caption{Optimal fit parameters for our model fits to RHESSI data.}
\label{paramstab}
\centering
\begin{tabular}{cccccccc}
\hline \hline
Flare & $f_0/S_0$ & $v_0$ & $T_e \, (\mathrm{MK})$ & $\delta$ &
$E_{\mathrm{fast}} \, (\mathrm{J})$ & $E_{\mathrm{therm}} \, (\mathrm{J})$ &
$E_{\mathrm{tot}} \, (\mathrm{J})$ \\
\hline
2002/03/17 & 1.46$\times 10^7$ & 1.18 & 21.2 & 6.30 & 3.79$\times
10^{25}$ & 1.43$\times 10^{24}$ & 3.94$\times 10^{25}$ \\
2002/08/21 & 3.40$\times 10^7$ & 2.66 & 29.7 & 4.86 & 1.47$\times
10^{23}$ & 1.38$\times 10^{23}$ & 2.84$\times 10^{23}$ \\
2002/08/22 & 5.28$\times 10^6$ & 2.90 & 25.0 & 5.69 & 3.75$\times 
10^{23}$ & 5.32$\times 10^{23}$ & 9.07$\times 10^{23}$ \\
\hline
\end{tabular}
\begin{list}{}{}
\item{$E_{\mathrm{fast}}$ and $E_{\mathrm{therm}}$ are the
  energy contents of the fast and background electrons respectively,
  calculated using the optimal parameters given.  $E_{\mathrm{tot}}$
  is the total energy content of all the electrons.}
\end{list}
\end{table}

\begin{table}
\caption{Optimal fit parameters for fits to RHESSI data using a simple
  thermal plus power-law model.}
\label{simpletab}
\centering
\begin{tabular}{cccccccc}
\hline \hline
Flare & $f_0/S_0$ & $v_0$ & $T_e \, (\mathrm{MK})$ & $\delta$ &
$E_{\mathrm{fast}} \, (\mathrm{J})$ & $E_{\mathrm{therm}} \, (\mathrm{J})$ &
$E_{\mathrm{tot}} \, (\mathrm{J})$ \\
\hline
2002/03/17 & 2.52$\times 10^8$ & 3.10 & 22.0 & 6.41 & 2.68$\times
10^{21}$ & 1.44$\times 10^{24}$ & 1.45$\times 10^{24}$ \\
2002/08/21 & 3.98$\times 10^8$ & 1.93 & 29.8 & 4.98 & 8.62$\times
10^{21}$ & 1.53$\times 10^{23}$ & 1.61$\times 10^{23}$ \\
2002/08/22 & 1.11$\times 10^7$ & 3.16 & 25.2 & 5.91 & 6.73$\times 
10^{22}$ & 5.37$\times 10^{23}$ & 6.05$\times 10^{23}$ \\
\hline
\end{tabular}
\begin{list}{}{}
\item{$E_{\mathrm{fast}}$ and $E_{\mathrm{therm}}$ are the
  energy contents of the fast and background electrons respectively,
  calculated using the optimal parameters given.  $E_{\mathrm{tot}}$
  is the total energy content of all the electrons.}
\end{list}
\end{table}

As previously discussed, our steady-state treatment is valid on
timescales longer than the collisional timescale but shorter than the
timescale for changes in the temperature of the flaring loop.  Thus, the
fitted model parameters describe the flare plasma at any instant, but
will change with time as the plasma evolves during the flare.  To
obtain a simple estimate of the energy content of the electrons in the
example flares, we fit a model spectrum to an observed RHESSI spectrum from
during the impulsive phase, and multiply the instantaneous energy
content by the duration of the phase.  To obtain the instantaneous energy, we insert the optimal fitted values of the relevant parameters into the source function, Eq.\ (\ref{Seq}), and calculate the total energy content of the fast electrons by integrating the electron kinetic energy over all possible electron velocities:
\begin{equation}
E_{\mathrm{fast}} \, = \, \int_{v_0}^{\infty} \frac{1}{2} m_e v^2 S(v)
v^2 {\,\mathrm{d}} v \, \mbox{.}
\label{Efasteq}
\end{equation}
The electron energy content of the thermal background plasma may also be
obtained from the optimal fitted parameters:
\begin{equation}
E_{\mathrm{therm}} \, = \, \frac{3}{2}NkT_e \, = \, \frac{3}{2} \left(
  f_0 v_{T}^2 \right) kT_e \, \mbox{.}
\label{Ethermeq}
\end{equation}
(For our present illustrative purposes, we assume a background plasma density of
$10^{15}$\,m$^{-3}$ -- typical of the lower corona -- to obtain the absolute value of $f_0$ from the
value of the emission measure as determined from the fits.)
The total energy in all the electrons is the sum of
$E_{\mathrm{fast}}$ and $E_{\mathrm{therm}}$.

We find single values of the source parameters to represent the whole
of the data time interval. The main assumption in doing this is that
background parameters ($f_0$, $T_e$) do not change. If this is the
case then interpreting the time integral of the data gives us the same
result as integrating a temporal sequence of data fits (as mentioned in
Section \ref{fp}; see also MacKinnon and Craig 1991). Although these parameters
may change, this will be partly because of the relaxation of the fast
electrons. Qualitatively, the procedure here may overestimate the
injected electron distribution, because temperature will increase, and
thus more of the observed photon spectrum will be due to `thermal'
electrons, as time goes on. To address this issue quantitatively we
would have to drop the linearisation of the Fokker-Planck equation,
Eq.~(\ref{fpeqn}), resulting in a considerably more complex problem that we do not
address here.  Such a fuller treatment would also allow us to precisely
determine the realm of validity of our linearisation.

The time intervals we use include the bulk of the hard X-ray emission from the
flares in question, but of course a more complete discussion of flare
energetics would also integrate over the entire history of the flare.

Also given in Table \ref{paramstab} are the total energy contents of
all electrons and of the injected and background electron populations 
individually for each event studied.  The flares of 21st and 22nd
August 2002 have comparable energies in the thermal and fast components,
and a total energy consistent with an M class flare.  Both also have
low-energy cutoffs at only a few times the thermal speed, indicating
that velocity diffusion will be relevant.  The optimal fitted parameters for
the flare of 17th March 2002 result in a much larger total energy which
would correspond to a larger and more energetic flare.
However, in this case the calculated energy content of the injected
electrons is many times greater than that of the background thermal
electrons.  This arises because the fitted low-energy cutoff is very
close to the thermal speed itself, and the power-law spectral index is
very large, meaning that the injected population will have a huge
number of electrons very near to the thermal speed.  These will
rapidly thermalise and give rise to the bulk of the Maxwellian portion of
the photon spectrum as shown in Fig.\ \ref{flareplots}, dominating the
emission from the background plasma.
  While this set of model
parameters corresponds to a minimum fit statistic, and produces a
model spectrum which closely reproduces the observations, they also
imply a situation where the injected electrons are no longer `dilute'
and our linearisation is no longer applicable.

\citet{Lin2003} employed Emslie's (2003) formulae for flare electron
energy content in their analysis of the 23rd July 2002 flare.  This
analysis assumed that the injected electrons had a power-law
low-energy cutoff at approximately the thermal speed ($T \approx
23$~MK) and, unlike our treatment, does not account for any
thermalisation of these lower energy injected electrons.  Lin et al.\
found the fast electron energy content for this X4.8 flare to be in
excess of $10^{27}$~J. This is very much greater than the highest total
energies ever deduced for the largest flares.  We also predict very
large energies in the injected electrons for cases where $v_0$ is very
low.  However, our estimates are not as extremely high as those made using
Emslie's formulae, since our treatment includes the appropriate
velocity diffusion effects for the lower energy injected electrons.

Due to the unambiguous nature of the straight portion of the spectral
profile at high energies, the fitted value of $\delta$ is
well-constrained.  However, as is evident from Figs.\ \ref{f0plot} and
\ref{v0plot}, variation of the values of $f_0/S_0$ and $v_0$ lead to similar
variations in the shape of the resulting spectral profile.  This
suggests the possibility of a degeneracy in the fitted values of these
parameters, which is indeed the case.  The quoted value of the fitted
$v_0$ given in the table is that for the most optimal fit.
However, it was found that the value of $v_0$ could be varied from
around one half to two times the optimal value for only a 10\%
increase in the value of the fit statistic.  Therefore we are
reluctantly forced to conclude that the value of $v_0$ is less
well-constrained
by the data than Emslie's original argument might suggest.
  Similarly, $f_0/S_0$, and to a lesser extent $T_e$, cannot
be unambiguously determined.
The value for the total electron energy varies by around an order of
magnitude when the value of $v_0$ is varied over our selected 10\%
range of fit statistic acceptability, with the total energy decreasing
as $v_0$ is increased.  Thus, while the model can reproduce
observed photon spectra, its nature may preclude a precise determination of the flare
electron energy content.

For comparative purposes, Table \ref{simpletab} gives fitting parameters as derived from fits using a
`simple' thermal plus power-law model, as may be employed in e.g.\ the OSPEX
package in the standard RHESSI analysis software
\citep[e.g.][]{schwartz}. As measured by our fit statistic, these fits are statistically acceptable at a similar
confidence level to our model fits.  As may be seen, the simple fits give
consistently larger values of $f_0/S_0$ (and comparable or slightly
greater temperatures) than our model fits, since all of the thermal
part of the spectrum must be accounted for by the Maxwellian
background with no contribution from thermalising fast electrons.
This results in higher energy contents for the thermal background
electrons, but
the total energy contents are lower than for our fits since the
thermalisation process is an energetically expensive way of producing
`thermal' photons.  The simple fit values for the low-energy cutoffs
are comparable with those of our model in that they are a few times
the thermal speed, but again this parameter is not well-constrained.
The largest difference is for the 17th March flare, for which our
model suggested an extremely low $v_0$.  The simple fit is more
conservative, and consequently does not give the very large energy
content in the injected electrons as derived from our model fits.

\section{Conclusions}
\label{end}
We have seen that for the case of accelerated flare electrons
impinging on a warm target, the effects of velocity diffusion should
not be neglected.  The process by which the lower energy electrons of
the injected population thermalise and merge with the ambient thermal
background is of importance when describing the behaviour of the electrons in this regime.  In particular, it has the effect of `smearing out' this region of the resulting bremsstrahlung photon spectrum, which in many cases is therefore not well described by a simple isothermal and power-law model.  However, this region can be modelled effectively by a treatment including velocity diffusion effects. 
 
A consequence of including velocity diffusion in the analysis is that
simple interpretation of observed photon spectra can be deceptive.
For example, as has been shown in Figs.\ \ref{f0plot}, \ref{deltaplot}
and \ref{v0plot}, determining the photon energy down to which the spectrum remains
power-law-like does not allow a simple evaluation of the parameters of the
injected electron population, most particularly its low energy cutoff.
 This in turn hinders determination of the flare electron energy content.

The surprising behaviour of the photon spectra is highlighted by the fits to RHESSI data in Fig.\
 \ref{flareplots}: visual inspection of the spectra would seem to suggest
 that the 17th March flare should have the highest temperature
  background plasma, since the Maxwellian portion of the spectrum is more
  prominent and extends to higher photon energy in this flare than in
  the others considered.  However, the fitted parameters imply that
  the 17th March flare actually has the lowest background plasma
  temperature.  The form and extent of the Maxwellian
  component of the photon spectrum in this case is completely
  dominated by thermalising electrons from the lower energies of the
  injected population.  This result emphasises that the thermalising
  process effectively couples the injected and the background
  populations, such that their contributions to the overall photon
  spectrum cannot easily be separated.  In effect, the distinction
  between the background and injected populations becomes rather
  arbitrary at these low energies, and it is no longer meaningful
  to distinguish between a background thermal electron and an
  electron which has thermalised out of the injected population.
  Further, because of this strong coupling it is also not possible to `swamp' the thermal
  region of the photon spectrum by contriving a large injected
  power-law population with a very low cutoff energy, as the lowest energy
  electrons will inevitably thermalise. However, this thermalisation
  process is an energetically expensive way to produce
  thermal emission.

The steady-state solution presented here has illuminated many of the consequences
of velocity diffusion in the context of solar flares.  However, an
approach which explicitly accounts for time-dependence would be an
interesting further development. This would allow modelling of the
evolution of the plasma parameters over the duration of a flare, and
may also be of benefit for more precise evaluations of the flare
energy budget.

\begin{acknowledgements}
We thank the referee, Dr.\ N.\ Vilmer, for helpful comments which improved the paper.
The work by P.H.\ was
funded by the U.K.\ Physical Sciences and Engineering Research
Council. R.K.G.\ is supported by a U.K.\ Particle Physics and
Astronomy Research Council CASE Award.
\end{acknowledgements}

\bibliographystyle{aa}

\begin{thebibliography}{16}
\expandafter\ifx\csname natexlab\endcsname\relax\def\natexlab#1{#1}\fi

\bibitem[{{Brown}(1971)}]{brown71}
{Brown}, J.~C. 1971, \solphys, 18, 489

\bibitem[{{Brown} {et~al.}(2003){Brown}, {Emslie}, \& {Kontar}}]{bro:al-03}
{Brown}, J.~C., {Emslie}, A.~G., \& {Kontar}, E.~P. 2003, \apjl, 595, L115

\bibitem[{{Emslie}(2003)}]{ems-03}
{Emslie}, A.~G. 2003, \apjl, 595, L119

\bibitem[{{Emslie} {et~al.}(2003)}]{Emslie_struct}  
{Emslie}, A.~G., {Kontar}, E.~P., {Krucker}, S., {Lin}, R.~P. 2003,
\apjl, 595, L107

\bibitem[{{Helander} \& {Sigmar}(2002)}]{hel_sig}
{Helander}, P., \& {Sigmar}, D.~J. 2002, {{C}ollisional {T}ransport in
{M}agnetized {P}lasmas} (Cambridge University Press)

\bibitem[{{Hoyng} {et~al.}(1976){Hoyng}, {Brown}, \& {van Beek}}]{hoy:al-76}
{Hoyng}, P., {Brown}, J.~C., \& {van Beek}, H.~F. 1976, \solphys, 48, 197

\bibitem[{{Kane} {et~al.}(1992){Kane}, {McTiernan}, {Loran}, {Fenimore},
  {Klebesadel}, \& {Laros}}]{kan:al-92}
{Kane}, S.~R., {McTiernan}, J., {Loran}, J., {Fenimore}, E.~E., {Klebesadel},
  R.~W., \& {Laros}, J.~G. 1992, \apj, 390, 687

\bibitem[{{Kontar} {et~al.}(2003){Kontar}, {Brown}, {Emslie}, {Schwartz},
  {Smith}, \& {Alexander}}]{Kontar2003}
{Kontar}, E.~P., {Brown}, J.~C., {Emslie}, A.~G., {Schwartz}, R.~A., {Smith},
  D.~M., \& {Alexander}, R.~C. 2003, \apjl, 595, L123

\bibitem[{{Lin} \& {Hudson}(1976)}]{linhud76}
{Lin}, R.~P., \& {Hudson}, H.~S. 1976, \solphys, 50, 153

\bibitem[{{Lin} {et~al.}(2002)}]{Lin2002}
{Lin}, R.~P., {et~al.} 2002, \solphys, 210, 3

\bibitem[{{Lin} {et~al.}(2003)}]{Lin2003}
{Lin}, R.~P., {et~al.} 2003, \apjl, 595, L69

\bibitem[{{MacKinnon} \& {Craig}(1991)}]{mac:cra-91}
{MacKinnon}, A.~L., \& {Craig}, I.~J.~D. 1991, \aap, 251, 693

\bibitem[{{McClements}(1987)}]{mcclements87}
{McClements}, K.~G. 1987, \aap, 175, 255

\bibitem[{{Montgomery} \& {Tidman}(1964)}]{mon:tid-64}
{Montgomery}, D.~C., \& {Tidman}, D.~A. 1964, {Plasma Kinetic Theory}
  (McGraw-Hill)

\bibitem[{{Piana} {et~al.}(2003){Piana}, {Massone}, {Kontar}, {Emslie},
  {Brown}, \& {Schwartz}}]{Piana2003}
{Piana}, M., {Massone}, A.~M., {Kontar}, E.~P., {Emslie}, A.~G., {Brown},
  J.~C., \& {Schwartz}, R.~A. 2003, \apjl, 595, L127

\bibitem[{{Press} {et~al.}(1992){Press}, {Teukolsky}, {Vetterling}, \&
  {Flannery}}]{num-rec}
{Press}, W.~H., {Teukolsky}, S.~A., {Vetterling}, W.~T., \& {Flannery}, B.~P.
  1992, {Numerical Recipes in C} (Cambridge University Press)

\bibitem[{{Rosenbluth} {et~al.}(1957){Rosenbluth}, {MacDonald}, \&
  {Judd}}]{ros:al-57}
{Rosenbluth}, M.~N., {MacDonald}, W.~M., \& {Judd}, D.~L. 1957, Phys.~Rev.,
  107, 1

\bibitem[{{Saint-Hilaire} \& {Benz}(2005)}]{saintbenz04}
{Saint-Hilaire}, P., \& {Benz}, A.~O. 2005, \aap, in press

\bibitem[{{Schwartz} {et~al.}(2002)}]{schwartz}
{Schwartz}, R.~A., {et~al.} 2002, \solphys, 210, 165S

\bibitem[{{Smith} {et~al.}(2002)}]{smith_2002}
{Smith}, D.~M., {et~al.} 2002, \solphys, 210, 33

\bibitem[{{Vilmer} \& {MacKinnon}(2003)}]{vilmac03}
{Vilmer}, N., \& {MacKinnon}, A.~L. 2003, in Energy Conversion and Particle
  Acceleration in the Solar Corona, ed. K.-L. {Klein} (Springer), 127--160

\end{thebibliography}

\end{document}